\RequirePackage[OT1]{fontenc}
\documentclass[letterpaper, 10 pt, conference]{ieeeconf}
\IEEEoverridecommandlockouts
\usepackage{amssymb, amsfonts, mathtools, xargs, tensor, units, cite, stmaryrd, mathrsfs, algpseudocode, algorithm, graphicx,subcaption,tikz,pgfplots, xcolor,cancel,multicol} 
\usepackage{tikz}
\usetikzlibrary{3d, calc}
\usepackage{pgfplots}
\usepgfplotslibrary{groupplots}
\pgfplotsset{compat=1.18}
\usepackage[unicode=true, colorlinks=false, hidelinks]{hyperref}

\definecolor{color1}{RGB}{0,128,0}      % Green
\definecolor{color2}{RGB}{255,0,0}      % Red
\definecolor{color3}{RGB}{0,0,255}      % Blue
\definecolor{color4}{RGB}{0,0,0}        % Black
\definecolor{color5}{RGB}{255,140,0}    % Dark Orange
\definecolor{color6}{RGB}{75,0,130}     % Indigo
\definecolor{color7}{RGB}{220,20,60}    % Crimson
\definecolor{color8}{RGB}{128,128,0}    % Olive
\definecolor{color9}{RGB}{0,139,139}    % Teal
\definecolor{color10}{RGB}{139,69,19}   % Brown
\definecolor{color11}{RGB}{128,0,128}   % Purple
\definecolor{color12}{RGB}{105,105,105} % Dim Gray

\def\BibTeX{{\rm B\kern-.05em{\sc i\kern-.025em b}\kern-.08em
    T\kern-.1667em\lower.7ex\hbox{E}\kern-.125emX}}
\begin{document}
\title{\LARGE \bf Decentralized Scalar Field Mapping using Gaussian Process}
\author{
Hossein Papi$^{1}$,  Muzaffar Qureshi$^{1}$, Kyle Volle$^{2}$, Rushikesh Kamalapurkar$^{1}$
\thanks{This research was supported in part by the Air Force Research Laboratory under contract number FA8651-24-1-0019. Any opinions, findings, or recommendations in this article are those of the author(s), and do not necessarily reflect the views of the sponsoring agencies.}% 
\thanks{$^{1}$ Department of Mechanical and Aerospace Engineering, University of Florida, email: {\tt\footnotesize \{hosseinpapi, muzaffar.qureshi, rkamalapurkar\} @ufl.edu}.}%
\thanks{$^2$ Torch Technologies, Shalimar, Florida, USA, email: {
\tt \footnotesize Kyle.Volle@torchtechnologies.com}
}}
\maketitle
\thispagestyle{empty}
\pagestyle{empty}

\begin{abstract}
Decentralized Gaussian process (GP) methods offer a scalable framework for multi-agent scalar-field estimation by replacing a centralized global model with multiple local models maintained by individual agents. A team of agents operates through overlapping domains; neighboring agents generally produce inconsistent distributions over shared regions. This paper investigates whether these inter-agent posterior discrepancies can be systematically exploited to improve team-level predictive performance and answers this question positively through a novel decentralized intersection data-sharing and assimilation protocol. Specifically, each agent constructs neighbor-specific packets from its local GP together with the geometry of the overlap between subdomains and selectively assimilates information received from neighboring agents to improve consistency of its posterior over the shared regions. The proposed architecture preserves locality in both computation and communication, supports decentralized neighbor-to-neighbor data assimilation, and allows local GP models to evolve cooperatively across the network without requiring the exchange full packet exchange or centralized inference.
\end{abstract}

\section{Introduction}
Mapping unknown scalar fields is important in diverse applications of autonomy, including surveillance, source localization, terrain assessment, and adaptive exploration \cite{SCC.Sabella.1998,SCC.Lin.Liu.ea2019,SCC.Matsuda.Nozaki.ea2023,SCC.Ren.Li.ea2021, SCC.Mao.Tan.ea2026}. Gaussian processes (GPs) have become a standard probabilistic tool for such mapping tasks because they provide a flexible nonparametric model together with principled uncertainty quantification \cite{SCC.Rasmussen.Williams2006,SCC.Qureshi.Ogri.ea2024}.

This problem naturally extends to decentralized multi-agent scenarios, where data are collected by multiple agents with limited communication, sensing range, and possibly inconsistent reference frames \cite{SCC.Ogri.Qureshi.ea2025a,qureshi2025gaussianprocessbasedscalarfield}. In the GP literature, communication-efficient inference schemes, such as product-of-experts models, Bayesian committee machines, federated learning, and decentralized information fusion, have been proposed to combine local GP models into a global posterior approximation while reducing computational and communication overhead \cite{muzaffar.Kontoudis.Stilwell2026}.

This paper considers a decentralized scalar-field estimation problem in which a global spatial domain is decomposed into local sensing subdomains assigned to multiple agents. Each agent acquires sparse, noisy measurements only within its own sensing region and constructs a local GP posterior using its locally available packets. Because these subdomains overlap, neighboring agents may simultaneously maintain posterior distributions over some of the same physical region while conditioning on different datasets. Even though the agents are modeling the same underlying scalar field, their posterior means and posterior uncertainties generally do not agree over the shared overlap regions.

These overlapping regions raise a fundamental, practically important question: can inter-agent discrepancies in posterior distributions over shared regions be exploited to improve the team's predictive performance? In standard decentralized estimation settings, disagreement between neighboring local models is often viewed primarily as an inconsistency to be reduced. In the setting studied here, however, such disagreement also carries information. A mismatch between neighboring posteriors indicates that the agents possess different local knowledge about the same portion of the field, and this mismatch may therefore provide a useful signal for deciding what information should be communicated, how it should be summarized, and how it should be assimilated by the receiving agent.

The main challenge is that this must be accomplished under decentralized constraints. No agent has direct access to the full network packets; each agent reasons only over its own local domain, and communication between agents should remain limited and localized. Consequently, the goal is not to reconstruct a centralized global GP posterior exactly, but rather to design a decentralized mechanism through which local GP models can evolve cooperatively by interacting over the overlap regions. Such a mechanism should improve consistency between neighboring predictions where their domains intersect, while preserving the locality of sensing, computation, and communication.

Motivated by this problem, this paper develops a framework for decentralized scalar-field estimation over decomposed domains with fixed overlap. In this framework, overlap regions are not treated as a secondary geometric feature, but as the key medium through which neighboring agents compare posterior behavior and exchange information relevant to local model refinement. The objective is to use posterior disagreement over shared regions in a principled way to improve both local and network-level predictive accuracy, while maintaining uncertainty-aware local GP representations and avoiding the exchange of full datasets or centralized inference.

\section{Problem Formulation}\label{sec:ProblemFormulation}
Given a compact domain $\mathcal{X} \subset \mathbb{R}^n$, let $f: \mathcal{X} \rightarrow \mathbb{R}$ denote an unknown scalar field representing the physical quantity of interest. Consider a network of $N$ mobile sensing agents indexed by $\mathcal{V}=\{1,\dots,N\}$, which communicate over a connected directed graph $\mathcal{G}=(\mathcal{V},\mathcal{E})$, where $\mathcal{E}\subseteq \mathcal{V}\times\mathcal{V}$ denotes the set of communication links. For each agent $i\in\mathcal{V}$, define the out-neighbor set as $\mathcal{N}_{i}^{\mathrm{out}}\coloneqq \{\, j\in\mathcal{V}\mid (i,j)\in\mathcal{E}\,\},$
and the in-neighbor set as $\mathcal{N}_{i}^{\mathrm{in}}\coloneqq \{\, j\in\mathcal{V}\mid (j,i)\in\mathcal{E}\,\}.$

To distribute the mapping task across the network, each agent $i \in \mathcal{V}$ is assigned a subdomain $\mathcal{X}^i \subseteq \mathcal{X}$ for measurements, and these subdomains collectively cover the domain, namely, $\mathcal{X} \subseteq \bigcup_{i\in\mathcal{V}} \mathcal{X}^i.$
For any pair of agents $i,j \in \mathcal{V}$, define the overlap region $\mathcal{O}^{ij} \coloneqq \mathcal{X}^i \cap \mathcal{X}^j.$

The agents evolve on a common synchronized measurement clock indexed by $t\in\mathbb{N}$. Let $\mathcal{D}_{t}^{i}\subseteq \mathcal{X}^{i}\times\mathbb{R}$ denote the set of measurements collected by agent $i$ up to clock step $t$. Each local measurement pair $(x_{\tau}^{i},y_{\tau}^{i})\in \mathcal{D}_{t}^{i}$ satisfies 
$y_{\tau}^{i}=f\!\left(x_{\tau}^{i}\right)+\epsilon_{\tau}^{i},$
where $x_{\tau}^{i}\in\mathcal{X}^{i}$ and $\epsilon_{\tau}^{i}\sim\mathcal{N}\!\left(0,(\sigma^{i})^{2}\right)$. The noise variables are assumed to be independent across agents and i.i.d. over $\tau$ for each fixed agent $i$. In addition to its local measurements \(\mathcal{D}_{t}^{i}\), agent \(i\) receives, from each in-neighbor \(j\in\mathcal{N}_{i}^{\mathrm{in}}\) a candidate library of edge-specific data packets. Each in-neighbor $j$ of agent $i$ uses their data $D_t^{j}$ and the overlap geometry $\mathcal{O}^{ji}$ to generate packets $\mathcal{C}_{t}^{j\to i}$ to transmit to agent $i$. Denote the rule used by $j$ to create the packet by $\Phi_{t}^{j\to i}$, i.e. $\mathcal{C}_{t}^{j\to i} = \Phi_{t}^{j\to i}\!\left(\mathcal{D}_{t}^{j},\mathcal{O}^{ji}\right)$. Each datum in $\mathcal{C}_{t}^{j\to i}$, referred to hereafter as a packet, has the form
\begin{equation}
\mathbf{u} = (u, m, s) \subseteq \mathcal{U} := \mathcal{X} \times \mathbb{R} \times \mathbb{R}{\geq 0},
\label{eq:packet_definition}
\end{equation}where \(u\) is a location, \(m\) is the sender's posterior mean at \(u\), and \(s\) is the sender's posterior variance at \(u\). The data received by agent \(i\) from all of its in-neighbors is referred to as the ``candidate packet library,'' defined as $\bar{\mathcal{C}}_{t}^{i} \coloneqq \bigcup_{j\in\mathcal{N}_{i}^{\mathrm{in}}} \mathcal{C}_{t}^{j\to i}.$ 

Agent \(i\) applies a local selection rule 
$\psi_{t}^{i}:\mathcal{P}(\mathfrak{U})\to \mathfrak{U},$
which maps the pooled candidate library to one selected packet to be assimilated into its local GP model. Let
\begin{equation}
\mathcal{M}_{t}^{i} \coloneqq\left\{\left(u_{\ell,t}^{i},m_{\ell,t}^{i},r_{\ell,t}^{i}\right) \right\}_{\ell=1}^{M_t^i} \subseteq \mathcal{X}\times\mathbb{R}\times\mathbb{R}_{>0}
\label{eq:fictitious_measurements_i}
\end{equation}
denote the set of packets selected for assimilation by agent \(i\) up to clock step \(t\). The local GP maintained by agent \(i\) is then conditioned on both its local measurements \(\mathcal{D}_{t}^{i}\) and the packets \(\mathcal{M}_{t}^{i}\), they have selected for assimilation from the data received from their neighbors. Defining the augmented information set $\mathcal{A}_{t}^{i} \coloneqq \left( \mathcal{D}_{t}^{i}, \mathcal{M}_{t}^{i} \right)$, the local GP of each agent is written as 
\begin{equation}
f \mid \mathcal{A}_{t}^{i} \sim \mathcal{GP}\!\left( \mu_{t}^{i}(\cdot), \Sigma_{t}^{i}(\cdot,\cdot) \right),\end{equation}
where \(\mu_{t}^{i}:\mathcal{X}\to\mathbb{R}\) is the posterior mean function and \(\Sigma_{t}^{i}:\mathcal{X}\times\mathcal{X}\to\mathbb{R}\) is the posterior covariance function. 

The objective in this paper is to test the hypothesis that the discrepancy between the local GPs on the overlaps of their sub-domains provides a useful decentralized metric that can be used to improve the predictive accuracy of the entire network. To that end, we design novel decentralized metrics, based on consistency of the local GPs on subdomain overlaps, that guide the design of the transmission rules $\Phi_t^{j\to i}$ and the selection rules $\psi_t^{i}$. We first describe the local GP models associated with the local measurements \(\mathcal{D}_t^i\), then introduce a sparse inducing-point approximation to generate reduced-order representations of the local GPs, and finally develop an inducing-point selection rule that determines which communication packets should be transmitted and assimilated by the agents.

\section{Methodology}
This section illustrates the local GP models maintained by each agent from its augmented information set. Later Section~\ref{sec:batch_targeted_ip_selection} describes the optimal selection of inducing points to construct targeted communication packets for neighboring agents.
\subsection{Gaussian Process Regression}
\label{sec:GP_section}
We first make precise the local GP model that each agent \(i\in\mathcal{V}\) generates from its augmented information set \(\mathcal{A}_t^i=(\mathcal{D}_t^i,\mathcal{M}_t^i)\). This GP model is the object that the transmission rule \(\Phi_t^{j\to i}\) and the selection rule \(\psi_t^i\) are ultimately intended to improve. The raw dataset \(\mathcal{D}_t^i\) contains locally collected measurements, while \(\mathcal{M}_t^i\) contains packets selected for assimilation from received packet libraries $\bar{\mathcal{C}}_{t}^{i}$. Each agent \(i\) employs a squared exponential covariance kernel of the form 
\begin{equation}
k^{i}(x,x') = \alpha^{i} \exp\!\left( -\frac{\|x-x'\|^{2}}{2(\ell^{i})^{2}} \right), \qquad x,x'\in\mathcal{X},\end{equation} where \(\alpha^{i}>0\) and \(\ell^{i}>0\) denote the signal variance and length-scale hyperparameters of agent \(i\), respectively.

Let $S_{t}^{i}\coloneqq \left|\mathcal{D}_{t}^{i}\right|, \  L_{t}^{i}\coloneqq \left|\mathcal{M}_{t}^{i}\right|,$
and write the local measurements of agent \(i\) at clock step \(t\) as $\mathcal{D}_{t}^{i} = \left\{
\left(x_{j,t}^{i},y_{j,t}^{i}\right)
\right\}_{j=1}^{S_{t}^{i}}$, where \(x_{j,t}^{i}\in\mathcal{X}^{i}\) and \(y_{j,t}^{i}\in\mathbb{R}\). Likewise, write the packets selected for assimilation as \eqref{eq:fictitious_measurements_i} where \(u_{\ell,t}^{i}\in\mathcal{X}\), \(m_{\ell,t}^{i}\in\mathbb{R}\), and \(r_{\ell,t}^{i}>0\) are the location, mean and variances estimates associated with the selected packet, respectively. For compactness of notation, define $X_{\mathrm{raw},t}^{i} \coloneqq \left[ x_{1,t}^{i},\dots,x_{R_t^i,t}^{i} \right]^{\top}, \qquad Y_{\mathrm{raw},t}^{i} \coloneqq \left[ y_{1,t}^{i},\dots,y_{L_t^i,t}^{i} \right]^{\top},$
$U_{t}^{i} \coloneqq \left[ u_{1,t}^{i},\dots,u_{R_t^i,t}^{i} \right]^{\top}, \qquad M_{t}^{i,\mathrm{fic}} \coloneqq \left[ m_{1,t}^{i},\dots,m_{L_t^i,t}^{i} \right]^{\top},$ and the augmented input and output arrays $X_{t}^{i} \coloneqq \begin{bmatrix} X_{\mathrm{raw},t}^{i}\\ U_t^i \end{bmatrix}, \qquad Y_{t}^{i} \coloneqq \begin{bmatrix} Y_{\mathrm{raw},t}^{i}\\ M_{t}^{i,\mathrm{fic}} \end{bmatrix}.$

Finally, define the associated diagonal noise matrix
\begin{equation}
R_t^i
\coloneqq
\operatorname{diag}
\left(
(\sigma^i)^2,\dots,(\sigma^i)^2, r_{1,t}^{i},\dots,r_{L_t^i,t}^{i}
\right).
\label{eq:augmented_noise_matrix}
\end{equation}
Conditioned on \(\mathcal{A}_{t}^{i}=(\mathcal{D}_t^i,\mathcal{M}_t^i)\), the local GP posterior of agent \(i\) at a test point \(x^{*}\in\mathcal{X}^{i}\) is given by $f(x^{*}) \mid \mathcal{A}_{t}^{i} \sim \mathcal{N}\!\left( \mu_{t}^{i}(x^{*}), \left(\sigma_{t}^{i}(x^{*})\right)^{2} \right),$ where \(\mu_{t}^{i}(x^{*})\) and \(\sigma_{t}^{i}(x^{*})\) denote the posterior mean and posterior standard deviation of agent \(i\), respectively.

Given the augmented information set \(\mathcal{A}_{t}^{i}\), define the kernel vector \(\mathbf{k}_{t}^{i}(x^{*})\in\mathbb{R}^{S_t^i+L_t^i}\) and kernel matrix \(K_{t}^{i}\in\mathbb{R}^{(S_t^i+L_t^i)\times(S_t^i+L_t^i)}\) by 
$\mathbf{k}_{t}^{i}(x^{*}) \coloneqq \left[ k^{i}(x^{*},X_t^i(1,:)), \dots, k^{i}(x^{*},X_t^i(R_t^i+L_t^i,:)) \right]^{\top},$
$\left[K_{t}^{i}\right]_{j,\ell} \coloneqq k^{i}(X_t^i(j,:),X_t^i(\ell,:)), \ j,\ell=1,\dots,R_t^i+L_t^i.$

Using \(\mathbf{k}_{t}^{i}(x^{*})\), \(K_{t}^{i}\), and the noise matrix \(R_t^i\) in \eqref{eq:augmented_noise_matrix}, the posterior mean and variance of agent \(i\) are computed as
\begin{equation}
\mu_{t}^{i}(x^{*}) = {\mathbf{k}_{t}^{i}(x^{*})}^{\top} \left( K_{t}^{i}+R_t^i \right)^{-1} Y_{t}^{i},
\end{equation} and
\begin{equation}\scalebox{0.9}{$
\left(\sigma_{t}^{i}(x^{*})\right)^{2} = k^{i}(x^{*},x^{*}) - {\mathbf{k}_{t}^{i}(x^{*})}^{\top} \left( K_{t}^{i}+R_t^i \right)^{-1} \mathbf{k}_{t}^{i}(x^{*}),$}
\end{equation}
where the first \(P_t^i\) diagonal entries of \(R_t^i\) correspond to local sensor-noise variance, and the remaining \(L_t^i\) diagonal entries correspond to the retained pseudo-measurement variances associated with previously selected packets.

Note that the computational cost of the GP prediction grows cubically with the total size of the augmented information set. In particular, since agent \(i\) conditions on \(D_t^i\) raw measurements and \(M_t^i\) retained fictitious measurements, each evaluation of the posterior mean requires \(\mathcal{O}\!\left((D_t^i+M_t^i)^3\right)\) time, and each evalaution of the posterior variance requires \(\mathcal{O}\!\left((D_t^i+M_t^i)^2\right)\) time. To reduce this burden while preserving the agent-specific local GP structure, the next subsection introduces an inducing-point approximation for each agent.

\subsection{Inducing Points in Gaussian Process}
\label{sec:inducing_points}
To obtain a scalable local approximation of a local GP, fix an arbitrary agent \(i\in\mathcal{V}\) and clock step \(t\in\mathbb{N}\). Recall from Section~\ref{sec:GP_section} that the local posterior of agent \(i\) is conditioned on the augmented information set \(\mathcal{A}_t^i=(\mathcal{D}_t^i,\mathcal{M}_t^i)\), whose total size is \(D_t^i+M_t^i\). To generate a reduced order representation of its GP, agent \(i\) employs a set of \(p_t^i\) inducing points, with \(p_t^i \ll D_t^i+M_t^i\) \cite{SCC.Rasmussen.Williams2006}.

Specifically, let 
\begin{equation}
\bar{X}_{t}^{i} \coloneqq \left[ \bar{x}_{1,t}^{i}, \dots, \bar{x}_{p_t^i,t}^{i} \right]^\top \subset \mathcal{X}^{i},\end{equation} denote the inducing-point set of agent \(i\) at clock step \(t\). Using the kernel \(k^i(\cdot,\cdot)\), define 
$\left[ K_{pp,t}^{i} \right]_{r,s} \coloneqq k^{i}(\bar{x}_{r,t}^{i}, \bar{x}_{s,t}^{i}), \ r,s = 1,\dots,p_t^i,$ and $\left[ K_{Xp,t}^{i} \right]_{j,r} \coloneqq k^{i}\!\left( X_t^i(j,:), \bar{x}_{r,t}^{i} \right), j = 1,\dots,M_t^i+L_t^i, \ r = 1,\dots,p_t^i,$ where \(X_t^i(j,:)\) denotes the \(j\)-th input location in the augmented input array \(X_t^i\), while \(\bar{x}_{r,t}^{i}\) denotes the \(r\)-th inducing point selected for the sparse approximation.

The corresponding low-rank approximation of the training covariance is
\begin{equation}
K_{t}^{i} \approx \tilde{K}_{t}^{i} \coloneqq K_{Xp,t}^{i} \left( K_{pp,t}^{i} \right)^{-1} \left( K_{Xp,t}^{i} \right)^\top.
\end{equation}
To equate the exact variance, we use the diagonal correction
\begin{equation}
\Sigma_{t}^{i,(p)}
\coloneqq
K_{Xp,t}^{i}
\left( K_{pp,t}^{i} \right)^{-1}
\left( K_{Xp,t}^{i} \right)^\top
+
\Lambda_{t}^{i,(p)}
+
R_t^i,
\label{eq:Sigma_inducing_i}
\end{equation}
where $\Lambda_{t}^{i,(p)} \coloneqq \operatorname{Diag} \left\{ K_{t}^{i} - K_{Xp,t}^{i} \left( K_{pp,t}^{i} \right)^{-1} \left( K_{Xp,t}^{i} \right)^\top \right\}.$ For a test point \(x^* \in \mathcal{X}^{i}\), define 
$k_{p,t}^{i}(x^{*})
\coloneqq
\left[
k^{i}(\bar{x}_{1,t}^{i},x^{*}),
\dots,
k^{i}(\bar{x}_{p_t^i,t}^{i},x^{*})
\right]^\top,$ 
$\Omega_{t}^{i}
\coloneqq
\Lambda_{t}^{i,(p)} + R_t^i,$ 
$\Gamma_{t}^{i}
\coloneqq
\left(\Omega_{t}^{i}\right)^{-1} K_{Xp,t}^{i},$
and $Q_{t}^{i}
\coloneqq
K_{pp,t}^{i}
+
\left(K_{Xp,t}^{i}\right)^\top
\left(\Omega_{t}^{i}\right)^{-1}
K_{Xp,t}^{i}.$
The resulting sparse GP predictive mean and variance are
\begin{equation}
\mu_{t}^{i,(p)}(x^{*})
=
\left(k_{p,t}^{i}(x^{*})\right)^\top
\left(Q_{t}^{i}\right)^{-1}
\left(\Gamma_{t}^{i}\right)^\top
Y_{t}^{i},
\label{eq:mu_inducing_i}
\end{equation}
\begin{multline}
\left(\sigma_{t}^{i,(p)}(x^{*})\right)^{2}
=
k^{i}(x^{*},x^{*})
\\
-
\left(k_{p,t}^{i}(x^{*})\right)^{\top}
\left[
\left(K_{pp,t}^{i}\right)^{-1}
-
\left(Q_{t}^{i}\right)^{-1}
\right]
k_{p,t}^{i}(x^{*}).
\label{eq:sigma_inducing_i}
\end{multline}
The inducing-point construction reduces the dominant local training cost from \(\mathcal{O}\!\left((D_t^i+M_t^i)^3\right)\) to \(\mathcal{O}\!\left((D_t^i+M_t^i)(p_t^i)^2 + (p_t^i)^3\right)\), and the storage cost from \(\mathcal{O}\!\left((D_t^i+M_t^i)^2\right)\) to \(\mathcal{O}\!\left((D_t^i+M_t^i)p_t^i + (p_t^i)^2\right)\). 

The computational burden in decentralized GP inference has two components: the \emph{local cost} incurred by each agent for training, storage, and prediction as its local augmented dataset evolves, and the \emph{communication cost} required to transmit candidate packets between agents. Because received information increases the computational load of the receiving agent and incurs a high communication cost, direct exchange of exact GP representations is not scalable. 

In this paper, we adopt the idea of inducing points to facilitate the design of the transmission rule $\Phi_{t}^{j\to i}$. Commensurate with the central hypothesis of the paper, the transmission rules are designed to reduce the mismatch between the local GPs on the overlapping subdomains. To that end, each sender selects a small set of inducing points for a specific receiver so that the transmitted packets are the most informative over their overlap region $\mathcal{O}^{ji}$. In the next subsection, an edge-dependent inducing-point selection rule is developed for communication-efficient inter-agent overlap consistency.

\section{Domain Targeted Inducing-Point Selection}
\label{sec:batch_targeted_ip_selection}
In Section~\ref{sec:ProblemFormulation}, the transmission rule \(\Phi_t^{j\to i}\) was introduced abstractly as a map from agent \(j\)'s retained information and the overlap geometry to a candidate set sent to agent \(i\). Here, we realize that idea by constructing edge-specific inducing points that are selected not to summarize the entirety of agent \(j\)'s subdomain, but to reduce predictive uncertainty over the portion of \(\mathcal{O}^{ji}\) that is most relevant to agent \(i\). Fix a communicating pair \((j,i)\in\mathcal{E}\) and a clock step \(t\in\mathbb{N}\). Let \(N_t^j \coloneqq D_t^j+M_t^j\) denote the total size of the augmented information set of sender agent \(j\). Let \(p_t^{j\to i}\in\mathbb{N}\) denote the total number of edge-specific inducing points to be selected by sender agent \(j\) for receiver agent \(i\) at clock step \(t\), and let $m\in\{0,1,\dots,p_t^{j\to i}-1\}$ denote the current greedy selection stage. The localized inducing-point method of \cite{Cole2021LIGP} selects inducing points for a single predictive location by minimizing a weighted integrated prediction-error criterion. Here, that construction is extended from a single target location to a finite target set, so that agent \(j\) selects edge-specific inducing points to improve prediction collectively over the portion of the overlap region that is most relevant to agent \(i\).

For notational convenience, write the squared exponential kernel of agent \(j\) as $k^{j}(x,x')
=
\nu^{j}\kappa^{j}(x,x'),
\nu^{j}\coloneqq(\alpha^{j})^{2}, $
$\kappa^{j}(x,x')
\coloneqq
\exp\!\left(
-\frac{\|x-x'\|^{2}}{\theta^{j}}
\right),
\ \theta^{j}\coloneqq 2(\ell^{j})^{2},$ where \(\alpha^{j}\) and \(\ell^{j}\) are the kernel hyperparameters already introduced in Section~\ref{sec:GP_section}.

Let $\mathcal{T}_{t}^{j\to i}
=
\left\{
z_{1,t}^{j\to i},\dots,z_{q_t^{j\to i},t}^{j\to i}
\right\}
\subseteq \mathcal{O}^{ji}$ denote the discretized target evaluation set in the overlap region, where \(q_t^{j\to i}\in\mathbb{N}\) is the number of target locations used for communication from agent \(j\) to agent \(i\) at clock step \(t\).  Let $\bar{X}_{m,t}^{j\to i}
=
\left[
\bar{x}_{1,t}^{j\to i},\dots,\bar{x}_{m,t}^{j\to i}
\right]^{\top}
\subseteq \mathcal{O}^{ji}$ be the current ordered edge-specific inducing-point set after \(m\) selections, and let $\bar{x}_{m+1,t}^{j\to i}
=
\left[
\bar{x}_{m+1,1,t}^{j\to i},\dots,\bar{x}_{m+1,n,t}^{j\to i}
\right]^{\top}
\in \mathcal{O}^{ji}$ denote a candidate next inducing point. The corresponding augmented inducing-point set is $\bar{X}_{m+1,t}^{j\to i}
=
\left[
\bar{x}_{1,t}^{j\to i},\dots,\bar{x}_{m,t}^{j\to i},\bar{x}_{m+1,t}^{j\to i}
\right]^{\top}.$

To evaluate the utility of \(\bar{x}_{m+1,t}^{j\to i}\) over the full target set \(\mathcal{T}_{t}^{j\to i}\), define the weighting function
\begin{equation}
\omega_{\mathcal{T}_{t}^{j\to i}}(\tilde{x})
=
\sum_{b=1}^{q_t^{j\to i}}
\alpha_{b,t}^{j\to i}\,
\kappa^{j}\!\left(\tilde{x},z_{b,t}^{j\to i}\right),
\quad
\sum_{b=1}^{q_t^{j\to i}}\alpha_{b,t}^{j\to i}=1,
\label{eq:domain_weight_function}
\end{equation}
where \(\alpha_{b,t}^{j\to i}\ge 0\) specifies the relative importance of the target location \(z_{b,t}^{j\to i}\).

Next, define the kernel vector $k_{m+1,t}^{j\to i}(x)
\coloneqq
\left[
k^{j}(x,\bar{x}_{1,t}^{j\to i}),
\dots,
k^{j}(x,\bar{x}_{m+1,t}^{j\to i})
\right]^{\top}
\in\mathbb{R}^{m+1},$ the Gram matrix $\left[K_{m+1,t}^{j\to i}\right]_{r,s}
\coloneqq
k^{j}(\bar{x}_{r,t}^{j\to i},\bar{x}_{s,t}^{j\to i}),
\qquad
r,s=1,\dots,m+1,$ and the cross-covariance matrix between the sender active inputs \(X_t^j=\{x_{u,t}^{j}\}_{u=1}^{M_t^j}\) and the augmented edge-specific inducing set $\left[K_{X,m+1,t}^{j\to i}\right]_{u,r}
\coloneqq
k^{j}(X_t^j(u,:),\bar{x}_{r,t}^{j\to i}),
\quad
u=1,\dots,N_t^j, 
r=1,\dots,m+1.$

Using the exact augmented training covariance matrix \(K_t^j\) from Section~\ref{sec:GP_section}, define $\scalebox{0.9}{$
\Lambda_{m+1,t}^{j\to i}
\coloneqq
\operatorname{Diag}
\left\{
K_t^j
-
K_{X,m+1,t}^{j\to i}
\left(K_{m+1,t}^{j\to i}\right)^{-1}
\left(K_{X,m+1,t}^{j\to i}\right)^{\top}
\right\},$}$ $\Omega_{m+1,t}^{j\to i}
\coloneqq
\Lambda_{m+1,t}^{j\to i}
+
R_t^j,$ and $Q_{m+1,t}^{j\to i}
\coloneqq
K_{m+1,t}^{j\to i}
+
\left(K_{X,m+1,t}^{j\to i}\right)^{\top}
\left(\Omega_{m+1,t}^{j\to i}\right)^{-1}
K_{X,m+1,t}^{j\to i}.$

The sparse predictive variance of sender agent \(j\), computed from the augmented information set \(\mathcal{A}_t^j\) and the augmented edge-specific inducing set \(\bar{X}_{m+1,t}^{j\to i}\), is then 
\begin{multline}
\left(\sigma_{m+1,t}^{j\to i}(\tilde{x})\right)^{2}
=
k^{j}(\tilde{x},\tilde{x})
-
\left(k_{m+1,t}^{j\to i}(\tilde{x})\right)^{\top} \\
\left[
\left(K_{m+1,t}^{j\to i}\right)^{-1}
-
\left(Q_{m+1,t}^{j\to i}\right)^{-1}
\right]
k_{m+1,t}^{j\to i}(\tilde{x}).\end{multline}

We now define the batch-targeted inducing-point criterion
\begin{multline}
\operatorname{BTIP}_{t}^{j\to i,(m+1)}
\!\left(
\bar{x}_{m+1,t}^{j\to i};
\mathcal{T}_{t}^{j\to i}
\right)
\\
=
\int_{\mathcal{O}^{ji}}
\omega_{\mathcal{T}_{t}^{j\to i}}(\tilde{x})\,
\frac{\left(\sigma_{m+1,t}^{j\to i}(\tilde{x})\right)^{2}}{\nu^{j}}
\,d\tilde{x}.
\label{eq:BTIP_integral}
\end{multline}
Since \(k^{j}(\tilde{x},\tilde{x})=\nu^{j}\), the normalization by \(\nu^{j}\) removes the process scale from the variance term. Although \eqref{eq:BTIP_integral} is induced by the predictive variance, we interpret it as a weighted spatially integrated surrogate of prediction error over the target region. Hence, minimizing this objective selects inducing points that reduce uncertainty near the prescribed target set \(\mathcal{T}_{t}^{j\to i}\).

Substituting \eqref{eq:domain_weight_function} into \eqref{eq:BTIP_integral} gives
\begin{multline}
\operatorname{BTIP}_{t}^{j\to i,(m+1)}
\!\left(
\bar{x}_{m+1,t}^{j\to i};
\mathcal{T}_{t}^{j\to i}
\right)
\\
=
\sum_{b=1}^{q_t^{j\to i}}
\alpha_{b,t}^{j\to i}
\int_{\mathcal{O}^{ji}}
\kappa^{j}\!\left(\tilde{x},z_{b,t}^{j\to i}\right)
\frac{\left(\sigma_{m+1,t}^{j\to i}(\tilde{x})\right)^{2}}{\nu^{j}}
\,d\tilde{x}.
\label{eq:BTIP_sum_form}
\end{multline}

For each target point \(z \in \mathcal{O}^{ji}\), define the matrix $W_{m+1,t}^{\ast,j\to i}(z) \in  \mathbb{R}^{(m+1)\times(m+1)}$ whose \((r,s)\)-entry is
\begin{equation}\scalebox{0.95}{$
\left[W_{m+1,t}^{\ast,j\to i}(z)\right]_{r,s}
\coloneqq
\frac{1}{\nu^{j}}
\int_{\mathcal{O}^{ji}}
\kappa^{j}(\tilde{x},z)\,
k^{j}(\tilde{x},\bar{x}_{r,t}^{j\to i})\,
k^{j}(\tilde{x},\bar{x}_{s,t}^{j\to i})
\,d\tilde{x},$}
\label{eq:BTIP_W_point}
\end{equation}
$r, s=1,\dots,m+1$. The corresponding batch-weighted matrix is $W_{m+1,t}^{\ast,j\to i}(\mathcal{T}_{t}^{j\to i})
=
\sum_{b=1}^{q_t^{j\to i}}
\alpha_{b,t}^{j\to i}\,
W_{m+1,t}^{\ast,j\to i}(z_{b,t}^{j\to i}).$

For \(\mathcal{O}^{ji}\), the criterion in \eqref{eq:BTIP_sum_form} admits the closed-form representation 
\begin{multline}
\operatorname{BTIP}_{t}^{j\to i,(m+1)}
\!\left(
\bar{x}_{m+1,t}^{j\to i};
\mathcal{T}_{t}^{j\to i}
\right) 
=\\
C_{\mathcal{T}_{t}^{j\to i}}
-
\operatorname{tr}\!\Bigg[
\Big(
(K_{m+1,t}^{j\to i})^{-1}
-
(Q_{m+1,t}^{j\to i})^{-1}
\Big)
W_{m+1,t}^{\ast,j\to i}(\mathcal{T}_{t}^{j\to i})
\Bigg],
\end{multline}
where
\begin{multline}
C_{\mathcal{T}_{t}^{j\to i}}
=
\sum_{b=1}^{q_t^{j\to i}}
\alpha_{b,t}^{j\to i}
\frac{\sqrt{\pi\theta^{j}}}{2}
\cdot\\
\prod_{\ell=1}^{n}
\left[
\operatorname{erf}\!\left(
\frac{z_{b,\ell,t}^{j\to i}-a_{\ell}^{j\to i}}{\sqrt{\theta^{j}}}
\right)
-
\operatorname{erf}\!\left(
\frac{z_{b,\ell,t}^{j\to i}-b_{\ell}^{j\to i}}{\sqrt{\theta^{j}}}
\right)
\right].
\end{multline}
where $a_{\ell}^{j\to i}$ and $b_{\ell}^{j\to i}$ denotes the overlap domain geometric constants. The next inducing point is selected greedily as $\bar{x}_{m+1,t}^{j\to i,\star}
=  
\arg\min_{\bar{x}\in\mathcal{O}^{ji}}
\operatorname{BTIP}_{t}^{j\to i,(m+1)}
\!\left(
\bar{x};
\mathcal{T}_{t}^{j\to i}
\right),$ and the edge-specific inducing-point set is updated according to $\bar{X}_{m+1,t}^{j\to i}  
= 
\left[
\bar{X}_{m,t}^{j\to i},
\bar{x}_{m+1,t}^{j\to i,\star}
\right].$

Thus, inducing points are selected sequentially to improve prediction over the target set \(\mathcal{T}_{t}^{j\to i}\).

For gradient-based optimization, the derivative of the objective with respect to coordinate \(\ell\in\{1,\dots,n\}\) of the candidate point \(\bar{x}_{m+1,t}^{j\to i}\) is
\begin{multline}
\frac{\partial}{\partial \bar{x}_{m+1,\ell,t}^{j\to i}}
\operatorname{BTIP}_{t}^{j\to i,(m+1)}
\!\left(
\bar{x}_{m+1,t}^{j\to i};
\mathcal{T}_{t}^{j\to i}
\right)
=
\\
-\operatorname{tr}\!\Bigg[
\Big(
\frac{\partial (K_{m+1,t}^{j\to i})^{-1}}{\partial \bar{x}_{m+1,\ell,t}^{j\to i}}
-
\frac{\partial (Q_{m+1,t}^{j\to i})^{-1}}{\partial \bar{x}_{m+1,\ell,t}^{j\to i}}
\Big)
W_{m+1,t}^{\ast,j\to i}(\mathcal{T}_{t}^{j\to i})
\Bigg]
\\
-\operatorname{tr}\!\Bigg[
\Big(
(K_{m+1,t}^{j\to i})^{-1}
-
(Q_{m+1,t}^{j\to i})^{-1}
\Big)
\frac{\partial W_{m+1,t}^{\ast,j\to i}(\mathcal{T}_{t}^{j\to i})}
{\partial \bar{x}_{m+1,\ell,t}^{j\to i}}
\Bigg],
\label{eq:BTIP_gradient}
\end{multline}
where
\begin{equation}
\frac{\partial W_{m+1,t}^{\ast,j\to i}(\mathcal{T}_{t}^{j\to i})}
{\partial \bar{x}_{m+1,\ell,t}^{j\to i}}
=
\sum_{b=1}^{q_t^{j\to i}}
\alpha_{b,t}^{j\to i}
\frac{\partial W_{m+1,t}^{\ast,j\to i}(z_{b,t}^{j\to i})}
{\partial \bar{x}_{m+1,\ell,t}^{j\to i}}.
\end{equation}
A practical initialization is given by the weighted centroid 
\begin{equation}
z_{c,t}^{j\to i}
=
\sum_{b=1}^{q_t^{j\to i}}
\alpha_{b,t}^{j\to i}\,
z_{b,t}^{j\to i},
\qquad
\bar{x}_{1,t}^{j\to i}=z_{c,t}^{j\to i}.
\end{equation}
After the edge-specific inducing-point set $\bar{X}_{t}^{j\to i,\star}
\coloneqq
\left[
\bar{x}_{1,t}^{j\to i,\star},\dots,\bar{x}_{p_t^{j\to i},t}^{j\to i,\star}
\right]^{\top}$ has been selected, sender agent \(j\) constructs the communicated packet library for receiver agent \(i\) by evaluating its local posterior at those locations. Specifically, the transmitted packet is 
\begin{multline}
\mathcal{C}_{t}^{j\to i}
= \Phi_t^{j \rightarrow i}\!\left(\mathcal{D}_t^{j}, \mathcal{O}^{ji}\right) \\
=
\left\{
\left(
\bar{x}_{r,t}^{j\to i,\star},
\mu_{t}^{j}\!\left(\bar{x}_{r,t}^{j\to i,\star}\right),
\Sigma_{t}^{j}\!\left(
\bar{x}_{r,t}^{j\to i,\star},
\bar{x}_{r,t}^{j\to i,\star}
\right)
\right)
\right\}_{r=1}^{p_t^{j\to i}}
\subseteq \mathfrak{U}.
\end{multline}
Thus, sender agent \(j\) does not transmit the entire local dataset or its full posterior representation. Instead, it transmits only the edge-specific inducing locations selected by the batch-targeted rule together with the sender posterior mean and variance predictions at those locations. These data packets provide a novel compact communication interface tailored to the overlap region \(\mathcal{O}^{ji}\), and they are the objects from which the receiver-side candidate library is formed in the next section.

\section{Receiver-Side Candidate Inducing-Point Selection}
\label{sec:receiver_global_one_point_selection}
At every measurement step \(t\), the receiver agent \(i \in \mathcal V\) has access to its local measurements \(\mathcal D_t^i\) and the edge-specific packets \(\mathcal{C}_t^{j \to i}\subseteq \mathfrak{U}\) received from all of its in-neighbors \(j \in \mathcal N_i^{\mathrm{in}}\). Given any candidate packet $\mathbf{u} = (u ,m ,s ) \in \bar{\mathcal{C}}_t^i$ as defined in \eqref{eq:packet_definition}, receiver agent \(i\) interprets \((m,s)\) as a one-point fictitious measurement at location \(u\). The corresponding updated posterior mean is
\begin{multline}
\mu_{i,t}^{+}(x;\mathbf{u}) = 
\mu_t^i(x)
\ + 
\Sigma_t^i(x,u)
\left(
\Sigma_t^i(u,u)+s
\right)^{-1}
\\
\times
\left(
m-\mu_t^i(u)
\right),
\label{eq:updated_receiver_mean_given_packet_short}
\end{multline}
and the corresponding updated posterior covariance is
\begin{multline}
\Sigma_{i,t}^{+}(x,x';\mathbf{u})
\ = 
\Sigma_t^i(x,x')
\ - 
\Sigma_t^i(x,u)
\left(
\Sigma_t^i(u,u)+s
\right)^{-1}
\\
\times
\Sigma_t^i(u,x').
\label{eq:updated_receiver_cov_given_packet_short}
\end{multline}

The receiver agent \(i\) selects one packet from the pooled candidate library \(\bar{\mathcal{C}}_t^i\) at each $t$. For any candidate packet \(\mathbf{u} \in \bar{\mathcal{C}}_t^i\), define the local receiver cost 
\begin{multline}
J_t^i(\mathbf{u})
\coloneqq
\|\sum_{j\in\mathcal{N}_i^{\mathrm{in}}}
\sum_{\mathbf{v}=(v,m_v,s_v)\in\mathcal{C}_t^{j\to i}}\\
\Big(
\alpha
\left|
\mu_{i,t}^{+}(v;\mathbf{u})-m_v
\right|^2 
+\;
\beta 
\left|
\Sigma_{i,t}^{+}(v,v;\mathbf{u})-s_v
\right|^2
\Big)\|
\end{multline} 
where $\alpha \ge 0$ and $\beta \ge 0$ are weights for mean and variance consistency, respectively. \(J_t^i(\mathbf{u})\) measures how well the receiver posterior obtained by assimilating packet \(\mathbf{u}\) agrees, at all received packet locations, with the GP posterior of its in-neighbors. The receiver-side packet selection rule for agent $i$ is therefore given as 
\begin{equation}
\mathbf{u}_t^{i,\star} = \psi_t^i(\bar{\mathcal{C}}_{t}^{i} )=
\arg\min_{\mathbf{u}\in\bar{\mathcal{C}}_t^i}
J_t^i(\mathbf{u}).
\label{eq:local_one_point_optimization_short}
\end{equation}
After the minimizing packet 
\begin{equation}
\mathbf{u}_t^{i,\star}=(u_t^{i,\star},m_t^{i,\star},s_t^{i,\star})
\end{equation}
is selected, receiver agent \(i\) treats it as a fictitious measurement with pseudo-noise variance 
$r_t^{i,\star} 
\coloneqq 
s_t^{i,\star}$. The packets selected for assimilation are then updated according to $\mathcal{M}_{t+1}^{i} 
= 
\mathcal{M}_{t}^{i} 
\cup 
\left\{
\left(
u_t^{i,\star},
m_t^{i,\star},
r_t^{i,\star}
\right)
\right\}.$
Thus, the selected packet is not used only for a temporary posterior correction at time \(t\), but is retained as a pseudo-measurement and incorporated into the augmented information set used to form the local GP posterior at subsequent measurement steps. Collecting the selected inducing points for all agents as
\begin{equation}
\mathbf{u}_t
\ \coloneqq 
\left(
\mathbf{u}_t^1,\dots,\mathbf{u}_t^N
\right)
\ \in 
\bar{\mathcal{C}}_t^1\times\cdots\times\bar{\mathcal{C}}_t^N.
\label{eq:global_decision_variable_short}
\end{equation}
and defining the network-level objective as
\begin{equation}
J_t(\mathbf{u}_t)
\ \coloneqq 
\sum_{i=1}^{N}
J_t^i(\mathbf{u}_t^i).
\label{eq:global_overlap_consistency_cost_short}
\end{equation}
Hence, the network-level packet-selection problem is
\begin{equation}
\mathbf{u}_t^\star \in \arg\min_{\mathbf{u}_t\in
\bar{\mathcal{C}}_t^1\times\cdots\times\bar{\mathcal{C}}_t^N}
J_t(\mathbf{u}_t).
\label{eq:global_one_point_optimization_short}
\end{equation}
Therefore, if each local problem has a unique minimizer, then the global problem also has a unique minimizer. In summary, the receiver-side packet-selection problem is a finite exact optimization that is globally defined but rigorously separable across receivers. Each receiver selects the packet that minimizes its local overlap-consistency cost, and the collection of these local minimizers simultaneously solves the network-level problem.

\section{Simulation Results}\label{sec:simulation}
The developed method was evaluated in a decentralized mapping simulation with $N=4$ over the square domain \([ -6,6 ] \times [ -6,6 ]\). The scalar field was generated from a smooth nonlinear test function over a \(41 \times 41\) evaluation grid. The four agent centers were placed at \((-3.0,-2.6)\), \((2.5,-2.8)\), \((3.0,2.8)\), and \((-2.7,3.0)\), and each agent was assigned a circular local domain of radius \(4.4\). At each clock step, every agent collected one local measurement, with a Gaussian noise standard deviation \(0.08\), over a total of \(20\) steps. 

For local GP modeling, the agent-specific length-scales are chosen as \([0.85,\ 1.10,\ 1.35,\ 0.95]\), the signal scales as \([0.90,\ 1.15,\ 1.05,\ 0.80]\), and the noise standard deviations as \([0.07,\ 0.09,\ 0.08,\ 0.10]\). At every step, each sender generated an edge-specific packet over each overlap region using \(4\) transmitted inducing points per directed edge. Each receiver then solved the exact finite-library packet-selection problem described in Section~\ref{sec:receiver_global_one_point_selection} and retained the selected packet as a fictitious measurement for subsequent updates. The retained pseudo-measurement variance was formed using the sender's predictive variance. In the receiver-side cost, the mean- and variance-consistency terms were weighted by \(\alpha=1.0\) and \(\beta=0.25\), respectively. 

\begin{figure}
\centering
\begin{tikzpicture}
\begin{axis}[
colormap/viridis,
width=0.82\linewidth,
axis equal image,
xmin=-6, xmax=6,
ymin=-6, ymax=6,
xlabel={$x$},
ylabel={$y$},
title={Global scalar field and four local agent domains},
colorbar,
point meta=explicit, scatter/use mapped color={draw=mapped color, fill=mapped color},
]
\addplot[
scatter,
only marks,
mark=square*,
mark size=1.6pt,
scatter src=explicit,
draw=none
]
table[x=x,y=y,meta=z,col sep=comma]{sim_data/truth_grid.csv};

\addplot[white,no marks,domain=0:360,samples=180,line width=0.9pt]
({-3.0 + 4.4*cos(x)},{-2.6 + 4.4*sin(x)});
\addplot[white,no marks,domain=0:360,samples=180,line width=0.9pt]
({ 2.5 + 4.4*cos(x)},{-2.8 + 4.4*sin(x)});
\addplot[white,no marks,domain=0:360,samples=180,line width=0.9pt]
({ 3.0 + 4.4*cos(x)},{ 2.8 + 4.4*sin(x)});
\addplot[white,no marks,domain=0:360,samples=180,line width=0.9pt]
({-2.7 + 4.4*cos(x)},{ 3.0 + 4.4*sin(x)});
\node[white,font=\bfseries] at (axis cs:-3.0,-2.6) {1};
\node[white,font=\bfseries] at (axis cs: 2.5,-2.8) {2};
\node[white,font=\bfseries] at (axis cs: 3.0, 2.8) {3};
\node[white,font=\bfseries] at (axis cs:-2.7, 3.0) {4};
\end{axis}
\end{tikzpicture}
\caption{True scalar field, local agent domains, and measurement locations.}
\label{fig:sim_truth_domains}
\end{figure}
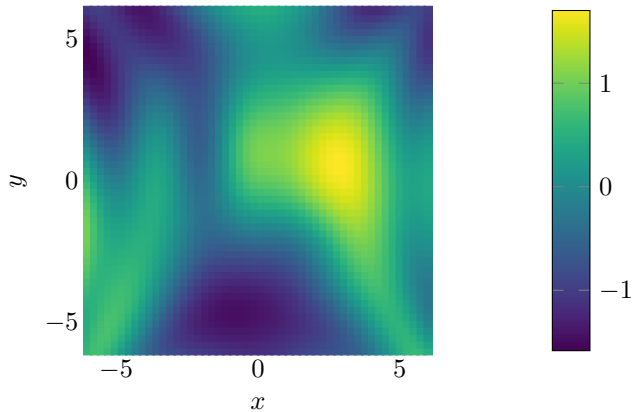

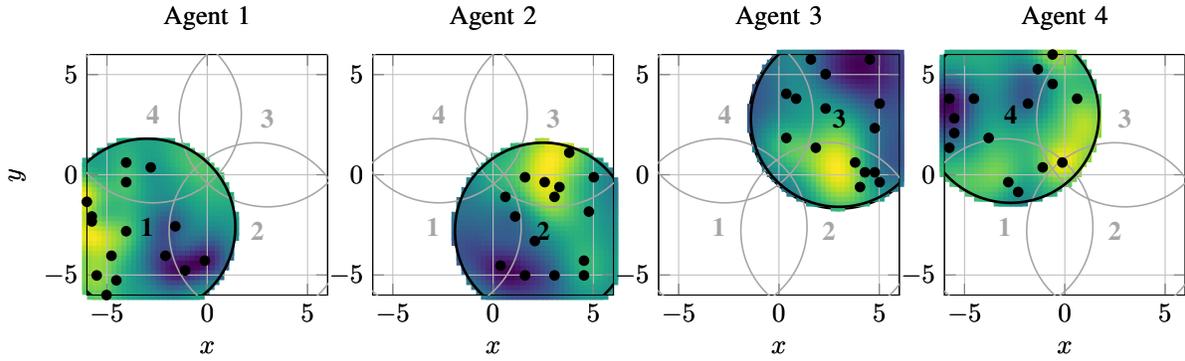
\begin{figure*}
\centering
\begin{tikzpicture}
\begin{groupplot}[
group style={group size=4 by 1, horizontal sep=0.6cm},
width=0.18\linewidth,
height=0.18\linewidth,
scale only axis,
xmin=-6, xmax=6,
ymin=-6, ymax=6,
xlabel={$x$},
colormap name=viridis,
point meta=explicit, scatter/use mapped color={draw=mapped color, fill=mapped color},
]

\nextgroupplot[title={Agent 1},ylabel={$y$},]
\addplot[
scatter,only marks,mark=square*,mark size=1.6pt,
scatter src=explicit,draw=none
]
table[x=x,y=y,meta=z,col sep=comma]{sim_data/final_mean_agent_1.csv};
\addplot[black,no marks,domain=0:360,samples=180,line width=0.8pt]
({-3.0 + 4.4*cos(x)},{-2.6 + 4.4*sin(x)});
\node[black,font=\bfseries] at (axis cs:-3.0,-2.6) {1};

\nextgroupplot[title={Agent 2}]
\addplot[
scatter,only marks,mark=square*,mark size=1.6pt,
scatter src=explicit,draw=none
]
table[x=x,y=y,meta=z,col sep=comma]{sim_data/final_mean_agent_2.csv};
\addplot[black,no marks,domain=0:360,samples=180,line width=0.8pt]
({2.5 + 4.4*cos(x)},{-2.8 + 4.4*sin(x)});
\node[black,font=\bfseries] at (axis cs:2.5,-2.8) {2};

\nextgroupplot[title={Agent 3}]
\addplot[
scatter,only marks,mark=square*,mark size=1.6pt,
scatter src=explicit,draw=none
]
table[x=x,y=y,meta=z,col sep=comma]{sim_data/final_mean_agent_3.csv};
\addplot[black,no marks,domain=0:360,samples=180,line width=0.8pt]
({3.0 + 4.5*cos(x)},{2.8 + 4.5*sin(x)});
\node[black,font=\bfseries] at (axis cs:3.0,2.8) {3};

\nextgroupplot[title={Agent 4}]
\addplot[
scatter,only marks,mark=square*,mark size=1.6pt,
scatter src=explicit,draw=none
]
table[x=x,y=y,meta=z,col sep=comma]{sim_data/final_mean_agent_4.csv};
\addplot[black,no marks,domain=0:360,samples=180,line width=0.8pt]
({-2.7 + 4.4*cos(x)},{3.0 + 4.4*sin(x)});
\node[black,font=\bfseries] at (axis cs:-2.7,3.0) {4};
\end{groupplot}
\begin{groupplot}[
group style={group size=4 by 1, horizontal sep=0.6cm},
width=0.18\linewidth,
height=0.18\linewidth,
scale only axis,
xmin=-6, xmax=6,
ymin=-6, ymax=6,
xlabel={$x$},
grid=both,
]
% ===================== Agent 1 =====================
\nextgroupplot[title={Agent 1},ylabel={$y$}]
\addplot[gray!70,no marks,domain=0:360,samples=180,line width=0.6pt]
({-3.0 + 4.4*cos(x)},{-2.6 + 4.4*sin(x)});
\addplot[gray!70,no marks,domain=0:360,samples=180,line width=0.6pt]
({ 2.5 + 4.4*cos(x)},{-2.8 + 4.4*sin(x)});
\addplot[gray!70,no marks,domain=0:360,samples=180,line width=0.6pt]
({ 3.0 + 4.4*cos(x)},{ 2.8 + 4.4*sin(x)});
\addplot[gray!70,no marks,domain=0:360,samples=180,line width=0.6pt]
({-2.7 + 4.4*cos(x)},{ 3.0 + 4.4*sin(x)});
\addplot[black,no marks,domain=0:360,samples=180,line width=1.0pt]
({-3.0 + 4.4*cos(x)},{-2.6 + 4.4*sin(x)});
\node[black,font=\bfseries] at (axis cs:-3.0,-2.6) {1};
\node[gray!70,font=\bfseries] at (axis cs: 2.5,-2.8) {2};
\node[gray!70,font=\bfseries] at (axis cs: 3.0, 2.8) {3};
\node[gray!70,font=\bfseries] at (axis cs:-2.7, 3.0) {4};
\addplot[
only marks,
mark=*,
mark size=1.8pt,
black,
fill=black,
]
table[x=x,y=y,col sep=comma]{sim_data/measurements_agent_1.csv};

% ===================== Agent 2 =====================
\nextgroupplot[title={Agent 2}]
\addplot[gray!70,no marks,domain=0:360,samples=180,line width=0.6pt]
({-3.0 + 4.4*cos(x)},{-2.6 + 4.4*sin(x)});
\addplot[gray!70,no marks,domain=0:360,samples=180,line width=0.6pt]
({ 2.5 + 4.4*cos(x)},{-2.8 + 4.4*sin(x)});
\addplot[gray!70,no marks,domain=0:360,samples=180,line width=0.6pt]
({ 3.0 + 4.4*cos(x)},{ 2.8 + 4.4*sin(x)});
\addplot[gray!70,no marks,domain=0:360,samples=180,line width=0.6pt]
({-2.7 + 4.4*cos(x)},{ 3.0 + 4.4*sin(x)});
\addplot[black,no marks,domain=0:360,samples=180,line width=1.0pt]
({ 2.5 + 4.4*cos(x)},{-2.8 + 4.4*sin(x)});
\node[gray!70,font=\bfseries] at (axis cs:-3.0,-2.6) {1};
\node[black,font=\bfseries] at (axis cs: 2.5,-2.8) {2};
\node[gray!70,font=\bfseries] at (axis cs: 3.0, 2.8) {3};
\node[gray!70,font=\bfseries] at (axis cs:-2.7, 3.0) {4};
\addplot[
only marks,
mark=*,
mark size=1.8pt,
black,
fill=black,
]
table[x=x,y=y,col sep=comma]{sim_data/measurements_agent_2.csv};

% ===================== Agent 3 =====================
\nextgroupplot[title={Agent 3}]
\addplot[gray!70,no marks,domain=0:360,samples=180,line width=0.6pt]
({-3.0 + 4.4*cos(x)},{-2.6 + 4.4*sin(x)});
\addplot[gray!70,no marks,domain=0:360,samples=180,line width=0.6pt]
({ 2.5 + 4.4*cos(x)},{-2.8 + 4.4*sin(x)});
\addplot[gray!70,no marks,domain=0:360,samples=180,line width=0.6pt]
({ 3.0 + 4.4*cos(x)},{ 2.8 + 4.4*sin(x)});
\addplot[gray!70,no marks,domain=0:360,samples=180,line width=0.6pt]
({-2.7 + 4.4*cos(x)},{ 3.0 + 4.4*sin(x)});
\addplot[black,no marks,domain=0:360,samples=180,line width=1.0pt]
({ 3.0 + 4.4*cos(x)},{ 2.8 + 4.4*sin(x)});
\node[gray!70,font=\bfseries] at (axis cs:-3.0,-2.6) {1};
\node[gray!70,font=\bfseries] at (axis cs: 2.5,-2.8) {2};
\node[black,font=\bfseries] at (axis cs: 3.0, 2.8) {3};
\node[gray!70,font=\bfseries] at (axis cs:-2.7, 3.0) {4};
\addplot[
only marks,
mark=*,
mark size=1.8pt,
black,
fill=black,
]
table[x=x,y=y,col sep=comma]{sim_data/measurements_agent_3.csv};

% ===================== Agent 4 =====================
\nextgroupplot[title={Agent 4}]
\addplot[gray!70,no marks,domain=0:360,samples=180,line width=0.6pt]
({-3.0 + 4.4*cos(x)},{-2.6 + 4.4*sin(x)});
\addplot[gray!70,no marks,domain=0:360,samples=180,line width=0.6pt]
({ 2.5 + 4.4*cos(x)},{-2.8 + 4.4*sin(x)});
\addplot[gray!70,no marks,domain=0:360,samples=180,line width=0.6pt]
({ 3.0 + 4.4*cos(x)},{ 2.8 + 4.4*sin(x)});
\addplot[gray!70,no marks,domain=0:360,samples=180,line width=0.6pt]
({-2.7 + 4.4*cos(x)},{ 3.0 + 4.4*sin(x)});
\addplot[black,no marks,domain=0:360,samples=180,line width=1.0pt]
({-2.7 + 4.4*cos(x)},{ 3.0 + 4.4*sin(x)});
\node[gray!70,font=\bfseries] at (axis cs:-3.0,-2.6) {1};
\node[gray!70,font=\bfseries] at (axis cs: 2.5,-2.8) {2};
\node[gray!70,font=\bfseries] at (axis cs: 3.0, 2.8) {3};
\node[black,font=\bfseries] at (axis cs:-2.7, 3.0) {4};
\addplot[
only marks,
mark=*,
mark size=1.8pt,
black,
fill=black,
]
table[x=x,y=y,col sep=comma]{sim_data/measurements_agent_4.csv};
\end{groupplot}
\end{tikzpicture}\vspace{-0.25cm}
\caption{Local GP prediction mean is shown at the culmination of the episode. The individual agent measurement locations are shown as black dots.}
\label{fig:sim_gp_means}
\end{figure*}

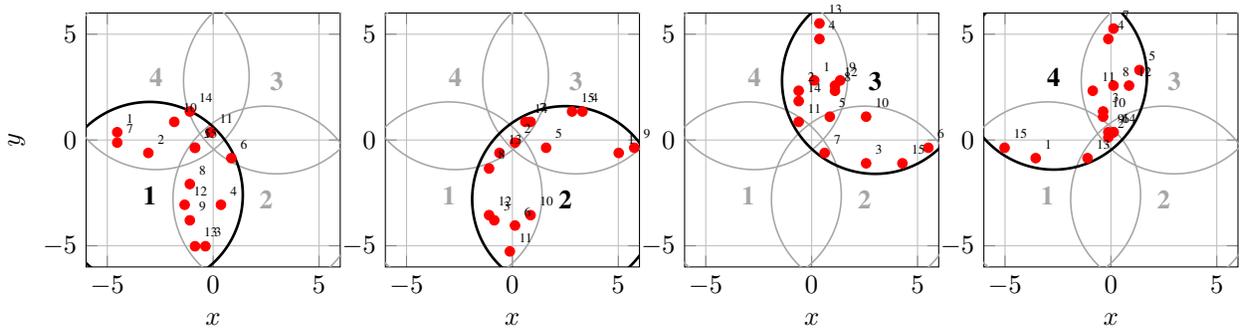
\begin{figure*}
\centering
\begin{tikzpicture}
\begin{groupplot}[
group style={group size=4 by 1, horizontal sep=0.6cm},
width=0.19\linewidth,
height=0.19\linewidth,
scale only axis,
xmin=-6, xmax=6,
ymin=-6, ymax=6,
xlabel={$x$},
grid=both,
]

% ===================== Agent 1 =====================
\nextgroupplot[ylabel={$y$},]
% all domains
\addplot[gray!70,no marks,domain=0:360,samples=180,line width=0.6pt]
({-3.0 + 4.4*cos(x)},{-2.6 + 4.4*sin(x)});
\addplot[gray!70,no marks,domain=0:360,samples=180,line width=0.6pt]
({ 2.5 + 4.4*cos(x)},{-2.8 + 4.4*sin(x)});
\addplot[gray!70,no marks,domain=0:360,samples=180,line width=0.6pt]
({ 3.0 + 4.4*cos(x)},{ 2.8 + 4.4*sin(x)});
\addplot[gray!70,no marks,domain=0:360,samples=180,line width=0.6pt]
({-2.7 + 4.4*cos(x)},{ 3.0 + 4.4*sin(x)});
% highlight agent 1
\addplot[black,no marks,domain=0:360,samples=180,line width=1.0pt]
({-3.0 + 4.4*cos(x)},{-2.6 + 4.4*sin(x)});
\node[black,font=\bfseries] at (axis cs:-3.0,-2.6) {1};
\node[gray!70,font=\bfseries] at (axis cs: 2.5,-2.8) {2};
\node[gray!70,font=\bfseries] at (axis cs: 3.0, 2.8) {3};
\node[gray!70,font=\bfseries] at (axis cs:-2.7, 3.0) {4};
% retained packets
\addplot[
only marks,
mark=*,
mark size=1.8pt,
red,
fill=red,
]
table[x=x,y=y,col sep=comma]{sim_data/retained_packets_agent_1.csv};
\addplot[
scatter,
only marks,
mark=none,
visualization depends on={value \thisrow{time} \as \pktlabel},
nodes near coords={\pktlabel},
nodes near coords style={font=\tiny, text=black, anchor=south west}
]
table[x=x,y=y,col sep=comma]{sim_data/retained_packets_agent_1.csv};

% ===================== Agent 2 =====================
\nextgroupplot[]
\addplot[gray!70,no marks,domain=0:360,samples=180,line width=0.6pt]
({-3.0 + 4.4*cos(x)},{-2.6 + 4.4*sin(x)});
\addplot[gray!70,no marks,domain=0:360,samples=180,line width=0.6pt]
({ 2.5 + 4.4*cos(x)},{-2.8 + 4.4*sin(x)});
\addplot[gray!70,no marks,domain=0:360,samples=180,line width=0.6pt]
({ 3.0 + 4.4*cos(x)},{ 2.8 + 4.4*sin(x)});
\addplot[gray!70,no marks,domain=0:360,samples=180,line width=0.6pt]
({-2.7 + 4.4*cos(x)},{ 3.0 + 4.4*sin(x)});
\addplot[black,no marks,domain=0:360,samples=180,line width=1.0pt]
({ 2.5 + 4.4*cos(x)},{-2.8 + 4.4*sin(x)});
\node[gray!70,font=\bfseries] at (axis cs:-3.0,-2.6) {1};
\node[black,font=\bfseries] at (axis cs: 2.5,-2.8) {2};
\node[gray!70,font=\bfseries] at (axis cs: 3.0, 2.8) {3};
\node[gray!70,font=\bfseries] at (axis cs:-2.7, 3.0) {4};
\addplot[
only marks,
mark=*,
mark size=1.8pt,
red,
fill=red,
]
table[x=x,y=y,col sep=comma]{sim_data/retained_packets_agent_2.csv};
\addplot[
scatter,
only marks,
mark=none,
visualization depends on={value \thisrow{time} \as \pktlabel},
nodes near coords={\pktlabel},
nodes near coords style={font=\tiny, text=black, anchor=south west}
]
table[x=x,y=y,col sep=comma]{sim_data/retained_packets_agent_2.csv};

% ===================== Agent 3 =====================
\nextgroupplot[]
\addplot[gray!70,no marks,domain=0:360,samples=180,line width=0.6pt]
({-3.0 + 4.4*cos(x)},{-2.6 + 4.4*sin(x)});
\addplot[gray!70,no marks,domain=0:360,samples=180,line width=0.6pt]
({ 2.5 + 4.4*cos(x)},{-2.8 + 4.4*sin(x)});
\addplot[gray!70,no marks,domain=0:360,samples=180,line width=0.6pt]
({ 3.0 + 4.4*cos(x)},{ 2.8 + 4.4*sin(x)});
\addplot[gray!70,no marks,domain=0:360,samples=180,line width=0.6pt]
({-2.7 + 4.4*cos(x)},{ 3.0 + 4.4*sin(x)});
\addplot[black,no marks,domain=0:360,samples=180,line width=1.0pt]
({ 3.0 + 4.4*cos(x)},{ 2.8 + 4.4*sin(x)});
\node[gray!70,font=\bfseries] at (axis cs:-3.0,-2.6) {1};
\node[gray!70,font=\bfseries] at (axis cs: 2.5,-2.8) {2};
\node[black,font=\bfseries] at (axis cs: 3.0, 2.8) {3};
\node[gray!70,font=\bfseries] at (axis cs:-2.7, 3.0) {4};
\addplot[
only marks,
mark=*,
mark size=1.8pt,
red,
fill=red,
]
table[x=x,y=y,col sep=comma]{sim_data/retained_packets_agent_3.csv};
\addplot[
scatter,
only marks,
mark=none,
visualization depends on={value \thisrow{time} \as \pktlabel},
nodes near coords={\pktlabel},
nodes near coords style={font=\tiny, text=black, anchor=south west}
]
table[x=x,y=y,col sep=comma]{sim_data/retained_packets_agent_3.csv};

% ===================== Agent 4 =====================
\nextgroupplot[]
\addplot[gray!70,no marks,domain=0:360,samples=180,line width=0.6pt]
({-3.0 + 4.4*cos(x)},{-2.6 + 4.4*sin(x)});
\addplot[gray!70,no marks,domain=0:360,samples=180,line width=0.6pt]
({ 2.5 + 4.4*cos(x)},{-2.8 + 4.4*sin(x)});
\addplot[gray!70,no marks,domain=0:360,samples=180,line width=0.6pt]
({ 3.0 + 4.4*cos(x)},{ 2.8 + 4.4*sin(x)});
\addplot[gray!70,no marks,domain=0:360,samples=180,line width=0.6pt]
({-2.7 + 4.4*cos(x)},{ 3.0 + 4.4*sin(x)});
\addplot[black,no marks,domain=0:360,samples=180,line width=1.0pt]
({-2.7 + 4.4*cos(x)},{ 3.0 + 4.4*sin(x)});
\node[gray!70,font=\bfseries] at (axis cs:-3.0,-2.6) {1};
\node[gray!70,font=\bfseries] at (axis cs: 2.5,-2.8) {2};
\node[gray!70,font=\bfseries] at (axis cs: 3.0, 2.8) {3};
\node[black,font=\bfseries] at (axis cs:-2.7, 3.0) {4};
\addplot[
only marks,
mark=*,
mark size=1.8pt,
red,
fill=red,
]
table[x=x,y=y,col sep=comma]{sim_data/retained_packets_agent_4.csv};
\addplot[
scatter,
only marks,
mark=none,
visualization depends on={value \thisrow{time} \as \pktlabel},
nodes near coords={\pktlabel},
nodes near coords style={font=\tiny, text=black, anchor=south west}]
table[x=x,y=y,col sep=comma]{sim_data/retained_packets_agent_4.csv};
\end{groupplot}
\end{tikzpicture}\vspace{-0.25cm}
\caption{Retained selected inducing-points for each agent. The number next to each packet denotes the time step.}
\label{fig:sim_packet_selection}
\end{figure*}

Figure~\ref{fig:sim_truth_domains} shows the true scalar field together with the four local agent domains. Figure~\ref{fig:sim_gp_means} shows the final local GP posterior mean for each agent at the end of the simulation, along with that agent’s measurement locations. Figure~\ref{fig:sim_packet_selection} shows the retained inducing-point packets for each agent, where the number next to each selected packet indicates the time step at which it was incorporated. 

\begin{figure*}
\centering
\begin{tikzpicture}
\begin{groupplot}[
group style={group size=3 by 1, horizontal sep=1.4cm},
width=0.19\linewidth,
height=0.19\linewidth,
scale only axis,
xlabel={Time step},
xmin=1,
xmax=15,
grid=both,
]

\nextgroupplot[
title={Network local RMSE},
ylabel={RMSE},
legend to name=sharedselflegend,
legend columns=2,
legend style={
font=\scriptsize,
draw=none,
 /tikz/every even column/.append style={column sep=0.3cm}
},
]
\addplot[blue,line width=1.5pt,mark=o,mark size=0.1pt]
table[x=t,y=shared_local_rmse,col sep=comma]{sim_data/shared_vs_self_metrics_history.csv};
\addlegendentry{Local dataset + induced points}
\addplot[red,dashed,line width=1.5pt,mark=square*,mark size=0.1pt]
table[x=t,y=self_local_rmse,col sep=comma]{sim_data/shared_vs_self_metrics_history.csv};
\addlegendentry{Local dataset only}

\nextgroupplot[
title={Network overlap RMSE},
ylabel={RMSE},
]
\addplot[blue,line width=1.5pt,mark=o,mark size=0.1pt]
table[x=t,y=shared_overlap_rmse,col sep=comma]{sim_data/shared_vs_self_metrics_history.csv};
\addplot[red,dashed,line width=1.5pt,mark=square*,mark size=0.1pt]
table[x=t,y=self_overlap_rmse,col sep=comma]{sim_data/shared_vs_self_metrics_history.csv};

\nextgroupplot[
title={Network local NLPD},
ylabel={NLPD},
]
\addplot[blue,line width=1.5pt,mark=o,mark size=0.1pt]
table[x=t,y=shared_nlpd,col sep=comma]{sim_data/shared_vs_self_metrics_history.csv};
\addplot[red,dashed,line width=1.5pt,mark=square*,mark size=0.1pt]
table[x=t,y=self_nlpd,col sep=comma]{sim_data/shared_vs_self_metrics_history.csv};

\end{groupplot}
\end{tikzpicture}

\vspace{0.15cm}
\pgfplotslegendfromname{sharedselflegend}
\vspace{-0.2cm}
\caption{Network-level comparison between the shared-information method and the self-only baseline averaged across agents}.
\label{fig:shared_vs_self_compare}
\end{figure*}
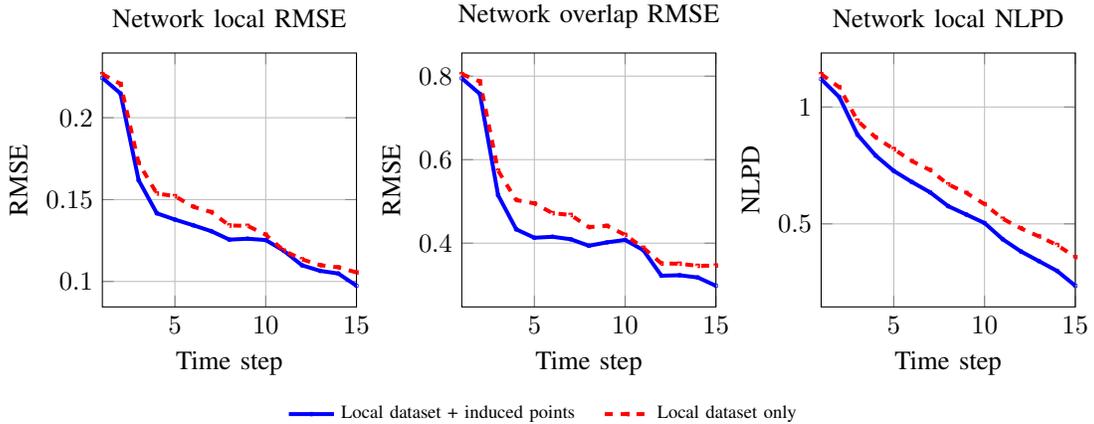

For an evaluation set \(\{x_n\}_{n=1}^{N}\), with ground-truth field value \(f(x_n)\), GP predictive mean \(\mu(x_n)\), and predictive variance \(\sigma^2(x_n)\), these metrics are defined as
\begin{equation}
\mathrm{RMSE}
=
\sqrt{\frac{1}{N}\sum_{n=1}^{N}\bigl(\mu(x_n)-f(x_n)\bigr)^2},
\end{equation}
\begin{equation}\scalebox{0.9}{$
\mathrm{NLPD}
=
\frac{1}{N}\sum_{n=1}^{N}
\left[
\frac{1}{2}\log\!\bigl(2\pi(\sigma^2(x_n)+\sigma_\epsilon^2)\bigr)
+
\frac{\bigl(f(x_n)-\mu(x_n)\bigr)^2}{2(\sigma^2(x_n)+\sigma_\epsilon^2)}
\right],$}
\end{equation}
where \(\sigma_\epsilon^2\) denotes the observation-noise variance. RMSE measures the posterior mean accuracy, while NLPD evaluates probabilistic calibration by jointly penalizing prediction error and uncertainty. Figure~\ref{fig:shared_vs_self_compare} compares the proposed inducing-point assimilation strategy against a self-only baseline at the network level. In the self-only baseline, each agent trains its GP using only locally collected measurements. In contrast, the proposed method augments each local dataset with selected inducing-point packets received from neighboring agents. This comparison highlights the impact of inter-agent communication on predictive performance within each agent’s local domain and in the overlap regions shared with other agents.

\section{Discussion}
\label{sec:discussion}
The results show that the developed method improves consistency over overlap regions while preserving accurate local GP predictions. Since each sender transmits only a compact overlap-targeted summary, the method remains decentralized and avoids sharing full local packets. A key feature is that communication has a cumulative effect. The selected packets are retained as fictitious measurements, so the local posteriors evolve using both newly collected measurements and previously accepted data packets. In comparison with GP models trained via local data sets, the developed algorithm performs better in the simulation studies, with results showing lower RMSE, along with reduced NLPD in the overlap regions while keeping the estimate errors preserved in the local map. This shows that agents improve their predictions and uncertainty estimates and reach better consensus than when operating in isolation.

The developed method currently assumes accurate transfer of information, but in practical scenarios, information sharing between agents can also introduce noise in the packets. This can degrade estimates, leading to reduced accuracy and increased error in the estimate compared to the local GP. Additionally, the framework relies on sufficient overlap between agents; with minimal overlap, the impact of information sharing diminishes and performance approaches that of isolated local models.

\section{Conclusion}
\label{sec:conclusion}
This paper developed a decentralized GP mapping method in which neighboring agents exchange compact edge-specific data packets to maintain coherent local posteriors over shared overlap regions. The proposed framework combines local GP modeling, batch-targeted sender-side packet construction, and an exact receiver-side finite-library selection rule that retains the best packet as a fictitious measurement.

The simulation results show that the method preserves local prediction accuracy while reducing overlap-region error through online communication. These results indicate that targeted packet exchange can improve posterior consistency in a decentralized GP network without sharing full local datasets. Future work will consider multi-packet receiver updates, adaptive hyperparameter learning, validation on real multi-robot sensing experiments, and development of a disturbance-tolerant distributed GP.

\small
\bibliographystyle{IEEETrans.bst}
\bibliography{scc,sccmaster,muzaffar,Hossein_LCSS}

\end{document}